\documentstyle[12pt,dina4,epsfig]{article}
\begin{document}

\title{Infrared Fixed Points for Ratios of Couplings in the Chiral
Lagrangian}

\author{Mario Atance\,\thanks{Work supported by the spanish FPU programme
under grant PF 97 25438231. E-mail atance@theo-physik.uni-kiel.de}\,
$^{(1)}$\ \ 
and\ \ 
Barbara Schrempp\,\thanks{E-mail schrempp@physik.uni-kiel.de}\,
$^{(2,1)}$ \\[9mm]
$^{(1)}${\small \it Institut f{\"u}r Theoretische Physik und Astrophysik, 
Universit{\"a}t Kiel, D-24118 Kiel, Germany} \\
$^{(2)}${\small \it Institut f{\"u}r Experimentelle und Angewandte Physik, 
Universit{\"a}t Kiel, D-24118 Kiel, Germany}}

\maketitle

\begin{abstract}
\noindent
In the framework of the low energy chiral Lagrangian the renormalization
group equations for the couplings are investigated up to order $p^6$ --
as well for SU(2)$\times$SU(2) as for SU(3)$\times$SU(3) chiral symmetry.
Infrared attractive fixed points for ratios of couplings are found. These
fixed point solutions turn out to agree with the values determined from 
experiment in a surprisingly large number of cases.

\smallskip
\noindent
PACS number(s): 11.10.Hi, 12.39.Fe

\end{abstract}

\bigskip

As is well known, an appropiate extension of QCD to low
energies is an effective field theory which realizes the spontaneously 
broken (approximate) chiral symmetry nonlinearly in terms of the light
Goldstone field degrees of freedom\cite{wei}. 
This symmetry information is encoded in the chiral Lagrangian. As in
any effective field theory, the Lagrangian has infinitely many 
contributing operators; however, they may be arranged according to their
importance for low energy observables in an expansion in powers of 
$p/\Lambda$. Here $p$ denotes the low momentum scale of interest and
$\Lambda$ some momentum cut-off, above which the chiral Lagrangian
ceases to be valid, with $p/\Lambda \leq 1$. The number of operators
contributing to each order in this expansion is finite. 
Similarly, perturbation theory in the number of loops and the renormalization
program can be carried out for effective field theories --- even though in 
principle infinitely many counterterms are required. The choice of a 
mass-independent renormalization scheme (MS, $\overline{{\mathrm MS}}$) 
leads, however, to counterterms which may again be arranged in an expansion
in powers of $p/\Lambda$. As a consequence, in any given order in $p/\Lambda$
the number of counterterms needed to absorb the divergences is again 
finite\cite{chi}.

Of interest are the coefficients of the operators in the chiral Lagrangian,
or rather --- after extraction of their dimension in form of powers of 
$\Lambda$ --- the dimensionless couplings. These couplings encode in 
principle the information about the QCD dynamics at higher scales.
Unfortunately, in practice they are unknown. There have been efforts to 
estimate some of them by using different techniques (lattice 
calculations, large $N_c$ limit, vector meson dominance, \ldots)\cite{chi2}.
The couplings in a given order of $p/\Lambda$ have to be determined by
experiment; one needs as many observables as couplings to be determined.
Within the framework of perturbation theory and renormalization described
above, the renormalization group equations for the couplings can be 
determined at any fixed order in $p/\Lambda$. They have all been calculated
up to and including $O((p/\Lambda)^4)$ --- abbreviated by $O(p^4)$ --- and
some of them\footnote{After
completion of this work, we became aware of the new reference 
\cite{nueva}. A more general investigation based on these impressive results
is in preparation.} up to $O(p^6)$\cite{p6}.

In this paper we investigate these renormalization group equations and 
search for infrared attractive fixed point solutions. This is a perfectly
legitimate search, since the infrared limit probes small momenta, where
the chiral Lagrangian is applicable. The analysis is performed as well for
SU(2)$\times$SU(2) as for SU(3)$\times$SU(3) chiral symmetry. As it turns
out, the renormalization group equations indeed exhibit infrared fixed points
in {\it ratios\/} of couplings. These fixed points are non-trivial special
solutions of the renormalization group equations which attract the 
renormalization group flow in its evolution from the scale $\Lambda$ towards
the infrared. It is very interesting to see, how they compare to the 
experimental values for the corresponding ratios of couplings (as far as
these are available). This comparison is performed.
Agreement is found for a surprisingly large number of ratios, thus confirming
to a certain extent the fixed point solutions by data.

A first reference to fixed points of the linearly realized chiral Lagrangian
was given in \cite{lin}. As the fixed point solutions correspond to 
renormalization
group invariant relations between couplings, they may also be viewed as 
solutions of the parameter reduction program \cite{oehme}. Earlier 
applications of the
parameter reduction technique in the framework of effective field theories
may be found in \cite{atance}.

\smallskip

The effective chiral Lagrangian for chiral $SU(3)\times SU(3)$
symmetry, expanded in terms of increasing powers of $p^2$, may be written as follows
\begin{eqnarray}
{\cal L}_\chi & = & {\cal L}^{(2)} + {\cal L}^{(4)} + {\cal L}^{(6)} + \cdots \nonumber \\
  & = & {\cal L}^{(2)} + \sum_i L_i {\cal O}_4^i + \sum_i {K_i\over \Lambda^2} 
        {\cal O}_6^i + \cdots 
\end{eqnarray}
where ${\cal O}_4$ and ${\cal O}_6$ are dimension four and six
operators, respectively, and where the dimensionful cut-off $\Lambda$
is introduced, leading to dimensionless couplings $L_i$ and $K_i$. The 
couplings to external right-handed and left-handed vector fields $r_\mu, l_\mu$, and
scalar and pseudoscalar fields $s, p$ are included.
The lowest order, $O(p^2)$, Lagrangian ${\cal L}^{(2)}$ may be written 
as
\begin{equation}
{\cal L}^{(2)} = {F_0^2\over 4} \langle D_\mu UD^\mu U^\dagger\rangle +
                 {F_0^2\over 4} \langle \chi U^\dagger + U\chi^\dagger\rangle.
\end{equation}
The operation
$\langle\cdot\rangle$ denotes the trace, the external fields enter through 
$\chi = 2B_0(s+ip)$ and the covariant derivative   
$D_\mu U = \partial_\mu U -ir_\mu U + iUl_\mu$. The unitary matrix $U$
\begin{equation}
U(\Phi) = \exp \left( i\sqrt{2} \Phi /F_0 \right)
\end{equation}
is given, for SU(3)$\times$SU(3), in terms of the Goldstone boson fields as follows
\begin{equation}
\Phi(x) \equiv \left( \begin{array}{ccc}
             {\pi^0\over \sqrt{2}}+{\eta\over\sqrt{6}} & \pi^+ & K^+ \\
             \pi^- & -{\pi^0\over \sqrt{2}}+{\eta\over\sqrt{6}} & K^0 \\
             K^- & \bar K^0 & -2{\eta\over\sqrt{6}}
                      \end{array} \right).
\end{equation}
${\cal L}^{(2)}$ involves two constants\cite{gass} $F_0, B_0$, which
are undetermined by the symmetry. They are related to the pion decay
constant and the quark condensate, respectively.

For the $O(p^4)$ Lagrangian ${\cal L}^{(4)}$ we follow the notation for
the couplings $L_i$ by Gasser and 
Leutwyler\cite{gass}. In the SU(3) case ${\cal L}_4$ involves
twelve terms:
\begin{eqnarray}
{\cal L}^{(4)} & = & \sum_{i=1}^{10} L_i {\cal O}_i + 
\sum_{i=1}^2 H_i {\cal O}'_i \\
 & = & L_1 \langle D_\mu U^\dagger D^\mu U\rangle^2 + 
       L_2 \langle D_\mu U^\dagger D_\nu U\rangle
           \langle D^\mu U^\dagger D^\nu U\rangle + 
       L_3 \langle D_\mu U^\dagger D^\mu U D_\nu U^\dagger D^\nu U\rangle\nonumber \\ 
  & & + L_4 \langle D_\mu U^\dagger D^\mu U\rangle 
         \langle U^\dagger\chi + \chi^\dagger U\rangle 
      + L_5 \langle D_\mu U^\dagger D^\mu U
         (U^\dagger\chi + \chi^\dagger U)\rangle 
      + L_6 \langle U^\dagger\chi + \chi^\dagger U\rangle^2 \nonumber \\  
 & &  + L_7 \langle U^\dagger\chi - \chi^\dagger U\rangle^2  
      + L_8 \langle U^\dagger\chi U^\dagger\chi + 
                 \chi^\dagger U\chi^\dagger U\rangle  
     -iL_9 \langle F_R^{\mu\nu}D_\mu U D_\nu U^\dagger + 
         F_L^{\mu\nu}D_\mu U^\dagger D_\nu U\rangle \nonumber \\
& &   +L_{10} \langle U^\dagger F_R^{\mu\nu}UF_{L\mu\nu}\rangle 
     +H_1 \langle F_{R\mu\nu}F_R^{\mu\nu} + F_{L\mu\nu}F_L^{\mu\nu}\rangle +
     H_2 \langle\chi^\dagger\chi\rangle \label{eq:lagrang}
\end{eqnarray}
where the field 
strenght tensors of the external gauge fields are
$F_{R\mu\nu} = \partial_\mu r_\nu - \partial_\nu r_\mu - i [r_\mu, r_\nu]$, 
and similarly for $l_\mu$.
The last two terms are not directly measurable, because they involve only the
external sources. Correspondingly we shall not consider them in the following. 
For the SU(2) case three couplings are redundant. 

The Lagrangian at order $p^6$ involves about hundred terms\cite{fea}, 
which we cannot write out in detail here.

Following the renormalization procedure for effective theories outlined
in the introduction, we use dimensional regularization and a mass
independent renormalization scheme. According to the renormalization program
for the chiral Lagrangian\cite{wei}, the
relations between bare and renormalized couplings are:
\begin{eqnarray}
L_i^b & = & \mu^{(d-4)} \left[ L_i^r + \Gamma_i \lambda \right] \label{1}\\
K_i^b & = & \mu^{2(d-4)} \left[ K_i^r + 
          \left( c_i + b_{ij} L_j^r \right) \lambda 
          + a_i\lambda^2 \right] ,\label{eq:doska}
\end{eqnarray}
where $\lambda = [(d-4)^{-1}-\zeta]/16\pi^2$,
and $\zeta$ is a constant which depends on the renormalization scheme. 
The constants $\Gamma_i$, $c_i$, and $b_{ij}$ are calculable; they
are scheme independent. 
It is now easy to obtain the renormalization 
group equations (RGE) for the renormalized couplings which have the
following generic form
\begin{eqnarray}
\mu{d L_i^r \over d\mu} & = & -{1\over 16\pi^2} \Gamma_i \label{Li}\\
\mu{d K_i^r \over d\mu} & = & -{1\over 16\pi^2} \left(c_i + b_{ij}L_j^r \right),\label{Ki}
\end{eqnarray}
where summation over repeated indices is implied. 
The values for the constants $\Gamma_i$ and for a selection of the
constants $c_i$ and $b_{ij}$ for $SU(2)$ and $SU(3)$
symmetry, respectively, will be given further down. Henceforth we
shall drop the index $r$, since we shall exclusively deal with
renormalized quantities. 

Let us next proceed with an analysis of the lowest order
renormalization group equation (\ref{Li}) for infrared (IR) fixed points
and their comparison to experiment. It is obvious that Eq.~(\ref{Li})
has no IR fixed point. However, the RGE for the ratios $L_i/L_j$ of
couplings 
\begin{equation}
\mu{d\over d\mu}\left({L_i\over L_j}\right)={1 \over 16\pi^2}{\Gamma_j \over L_j}
\left({L_i\over L_j} - {\Gamma_i\over \Gamma_j}\right) 
\end{equation}
has a fixed point at
\begin{equation}
\left. {L_i \over L_j} \right|_{{\mathrm f.p.}}  = {\Gamma_i \over
\Gamma_j}\ \ \ {\rm for}\ \ \ \Gamma_i,\ \Gamma_j\neq 0. \label{eq:ls}
\end{equation}
The general solution of the RGE (\ref{Li}) for initial value
$L(\Lambda)$ at $\mu=\Lambda$ is.
\begin{equation}
L_i(\mu)=L_i(\Lambda)-{1\over 16\pi^2} \Gamma_i\;log{\mu\over \Lambda}
\end{equation}
One
reads off that the approach of the ratio of $L_i/L_j$ of general
solutions $L_i,\ L_j$ to the IR fixed
point is controlled by the {\it fast} approach of $log(\mu/
\Lambda)\rightarrow -\infty$ in the IR limit $\mu\rightarrow 0$. 

It is now interesting to compare the experimental values for the
ratios of couplings with the predictions obtained 
for their fixed points (\ref{eq:ls}).

The following tables summarize the experimental values of the
couplings for $SU(2)$ symmetry\footnote{The couplings $l_i$ are defined
as in Refs.~\cite{gass,cou}. They are not identical with those given in
Eq.~(\ref{eq:lagrang}).} \cite{cou,gass} at an energy
scale of the order of the pion mass and for $SU(3)$ symmetry
\cite{3cou} at an energy
scale $\mu = m_\rho$ (which is unfortunately rather large). These values are
derived from meson
decay constants, electromagnetic form factors and, for $SU(3)$ symmetry, 
also from semileptonic kaon decays. Also the corresponding  
coefficients $\Gamma_i$ of the beta functions are given.
\begin{center}
$SU(2)$:\ \ \begin{tabular}{|c|cccccc|} \hline
i & 1 & 2 & 3 & 4 & 5 & 6  \\ \hline
$l_i\cdot 10^3$ & $-2.4\pm 3.9$ & $12.7\pm 2.7$ & $-4.6\pm 3.8$ 
& $27.2\pm 5.7$ & $-7.3\pm 0.7$ & $17.4\pm 1.2$ \\ 
$\Gamma_i$ & 1/3 & 2/3 & -1/2 & 2 & -1/6 & -1/3 \\ \hline 
\end{tabular}
\end{center}
\begin{center}
$SU(3)$:\ \ \begin{tabular}{|c|ccccc|} \hline
i & 1 & 2 & 3 & 4 & 5 \\ \hline
$L_i\cdot 10^3$ & $0.4\pm 0.3$ & $1.35\pm 0.3$ & $-3.5\pm 1.1$ & 
$-0.3\pm 0.5$ & $1.4\pm 0.5$ \\ 
$\Gamma_i$ & 3/32 & 3/16 & 0 & 1/8 & 3/8\\ \hline
\end{tabular}
\end{center}
\hspace*{2cm}\begin{tabular}{|c|ccccc|} \hline
i & 6 & 7 & 8 & 9 & 10\\ \hline
$L_i\cdot 10^3$& $-0.2\pm 0.3$ & $-0.4\pm 0.2$ & $0.9\pm 0.3$ & 
$6.9\pm 0.7$ & $-5.5\pm 0.7$ \\ 
$\Gamma_i$& 11/144 & 0 & 5/48 & 
1/4 & -1/4 \\ \hline
\end{tabular}

\begin{figure}[ht]
\vspace{-1.0cm}
\begin{center}
\leavevmode
\epsfxsize=12cm
\epsfbox{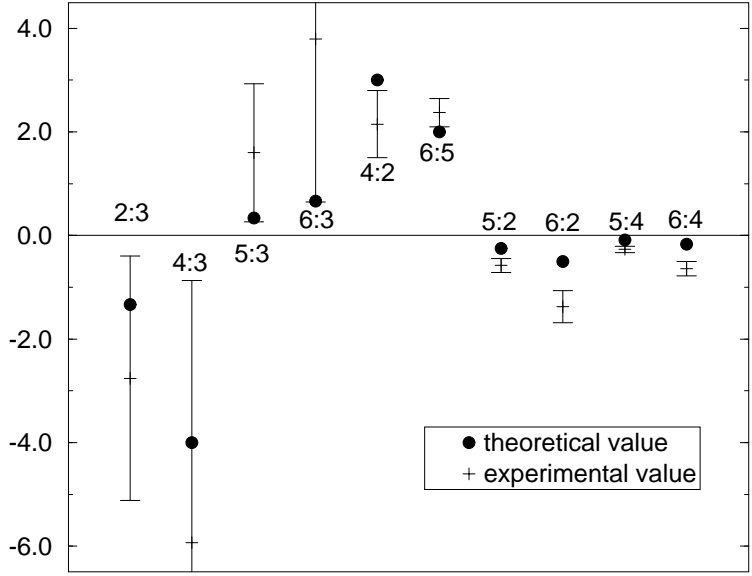}
\end{center}
\vspace{-2.0cm}
\caption{Ratios $L_i/L_j$ of the experimental values of the couplings
$L_i$ in comparison with the corresponding IR fixed point ratios
$\Gamma_i/\Gamma_j$ for $SU(2)$ symmetry. To guide the eye, the ratios 
are ordered from left to right according to decreasing agreement.} \label{fig:su2}
\end{figure}

\begin{figure}[ht]
\vspace{-1.0cm}
\begin{center}
\leavevmode
\epsfxsize=12cm
\epsfbox{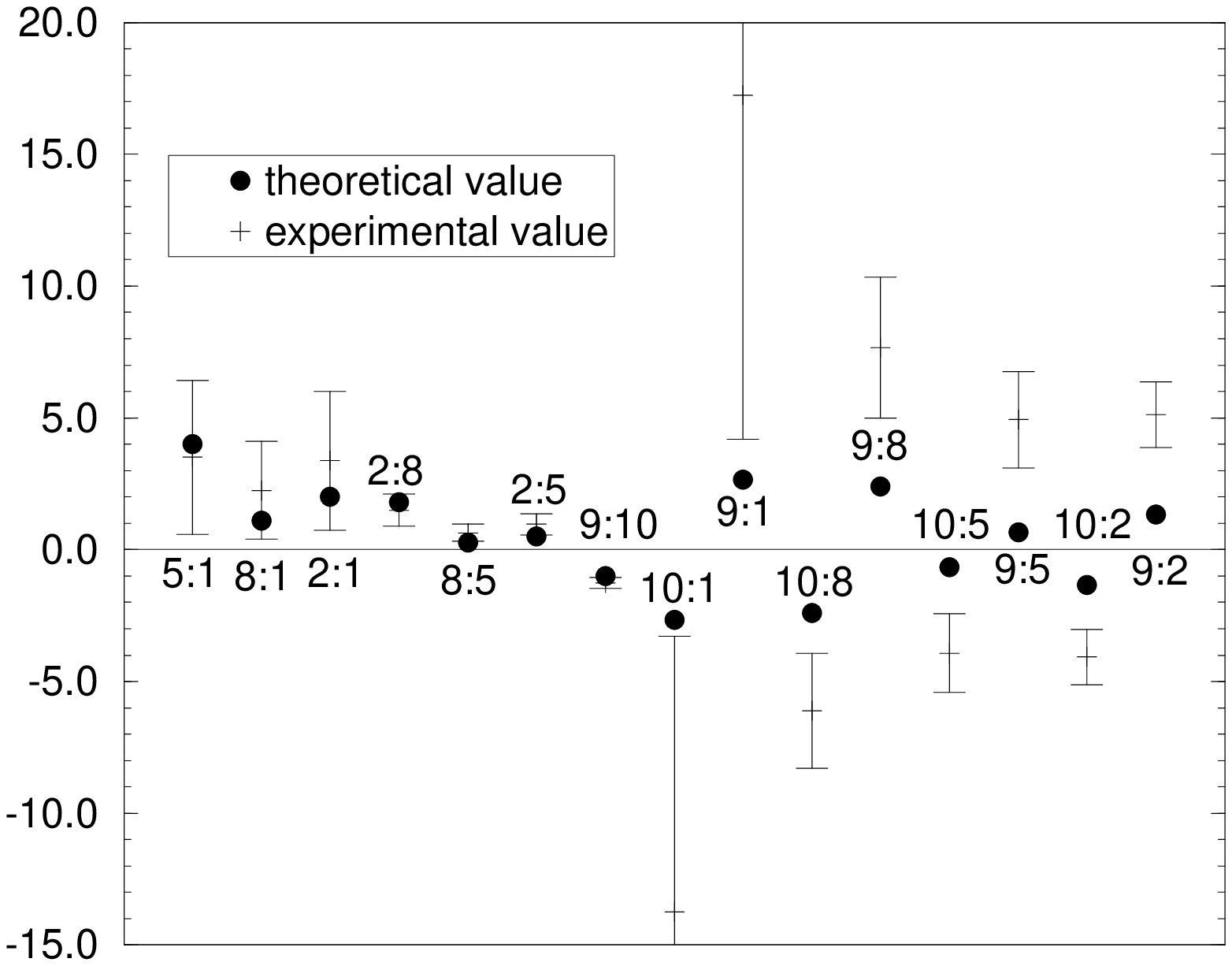}
\end{center}
\vspace{-2.0cm}
\caption{As in Fig.\,\ref{fig:su2}, but for $SU(3)$.} \label{fig:su3}
\end{figure}

\smallskip

In Figs. \ref{fig:su2} and  \ref{fig:su3} the experimental ratios
$L_i/L_j$ are compared  with their fixed point ratios
$\Gamma_i/\Gamma_j$ for values of $i,\ j$ for which $\Gamma_i$ and
$\Gamma_j$ are nonzero. In addition we omitted  $L_1$ for $SU(2)$  and 
$L_4$ and $L_6$ for $SU(3)$, since their errors are too large to give
meaningful comparisons. To guide the eye, we ordered the ratios from
left to right with decreasing agreement of experimental values with
the fixed point values. The overall agreement between experimental and
predicted values is quite impressive. Even, where the agreement fails
on a quantitative level (at the right end of the figures), the sign and
the qualitative tendency are correct and the deviation is generically of
the order of one standard deviation.  

Next we proceed to the analysis of the $O(p^6)$ RGE~(\ref{Ki}). We
calculated the renormalization group equations for a selection of
couplings $K_i$, based on results from Ref.\,\cite{bij} for $SU(2)$
symmetry and from Ref.\,\cite{gol} for $SU(3)$ symmetry. We use the 
notation of Refs.\,\cite{fea,gol}.

For $SU(2)$ symmetry the RGE, for a subset of the couplings, are:\\[-12mm]

\begin{eqnarray}
\mu{dK_1 \over d\mu} & = & {1\over (16\pi^2)^2}{193\over 27} + 
{1\over 16\pi^2}\left( {208\over 3}L_1 +{112\over 3}L_2 +24L_3+2L4 \right)\nonumber \\
\mu{dK_2 \over d\mu} & = & -{1\over (16\pi^2)^2}{556\over 27} - 
{1\over 16\pi^2}\left( 136L_1 +{248\over 3}L_2 +28L_3+2L4 \right)\nonumber \\
\mu{dK_3 \over d\mu} & = & {1\over (16\pi^2)^2}{755\over 108} + 
{1\over 16\pi^2}\left( {200\over 3}L_1 +44 L_2 \right)\nonumber \\
\mu{dK_4 \over d\mu} & = & -{1\over (16\pi^2)^2}{1\over 108} - 
{1\over 16\pi^2}\left( {4\over 3}L_1 +{4\over 3} L_2 \right)\nonumber \\
\mu{dK_5 \over d\mu} & = & -{1\over (16\pi^2)^2}{29\over 432} - 
{1\over 16\pi^2}\left( {21\over 2}L_1 +{107\over 12} L_2 \right)\nonumber \\
\mu{dK_6 \over d\mu} & = & -{1\over (16\pi^2)^2}{79\over 432} - 
{1\over 16\pi^2}\left( {5\over 6}L_1 +{25\over 12} L_2 \right) 
\end{eqnarray}

And for $SU(3)$ symmetry we consider the RGE:\\[-12mm]

\begin{eqnarray}
\mu{dK\over d\mu} & = & -{1\over (16\pi^2)^2}{43\over 96} -
{1\over 16\pi^2}\left( {104\over 9} L_1+{26\over 9}L_2+
{61\over 18}L_3-{34\over 9}L_4+L_5-4L_6-2L_8 \right) \nonumber\\
\mu{dA\over d\mu} & = & {1\over (16\pi^2)^2}{175\over 288} +
{1\over 16\pi^2}\left( {28\over 3} L_1+{34\over 3}L_2+
{25\over 3}L_3-{26\over 3}L_4+{8\over 3}L_5+12L_6
-12L_8 \right) \nonumber\\ 
\mu{dB\over d\mu} & = & -{1\over (16\pi^2)^2}{19\over 48} +
{1\over 16\pi^2}\left( {32\over 9}L_1+{8\over 9}L_2+
{8\over 9}L_3-{106\over 9}L_4+{22\over 9}L_5+20L_6 \right) \nonumber\\ 
\mu{dC\over d\mu} & = & -{1\over (16\pi^2)^2}{691\over 2592} -
{1\over 16\pi^2}\left( {28\over 9} L_1+{34\over 9}L_2+
{59\over 18}L_3-{26\over 9}L_4+3L_5+4L_6-6L_8 \right) \nonumber \\ 
\mu{dD\over d\mu} & = & -{1\over (16\pi^2)^2}{20\over 27} +
{1\over 16\pi^2}\left( {20\over 3} L_4+{23\over 3}L_5-
{40\over 3}L_6-40L_7-{86\over 3}L_8 \right) \nonumber\\ 
\mu{dE\over d\mu} & = & {1\over (16\pi^2)^2}{5\over 162} +
{1\over 16\pi^2}\left( {44\over 9} L_4+{1\over 27}L_5-
{88\over 9}L_6-+{136\over 9}L_7-{134\over 27}L_8 \right) \nonumber\\ 
\mu{dF\over d\mu} & = & {1\over (16\pi^2)^2}{167\over 144} -
{1\over 16\pi^2}\left( 8L_6-64L_7-{62\over 3}L_8 \right) \nonumber\\ 
\mu{dG\over d\mu} & = & -{1\over (16\pi^2)^2}{371\over 648} -
{1\over 16\pi^2}\left( 2L_5-{16\over 3}L_6+72L_7+24L_8 \right) \nonumber\\ 
\mu{dH\over d\mu} & = & {1\over (16\pi^2)^2}{9\over 16} +
{1\over 16\pi^2}\left( L_5+{152\over 3}L_7+{62\over 3}L_8 \right) 
\end{eqnarray}

where we defined, as in Ref.\,\cite{gol}, the following combinations of 
couplings:
\begin{equation}
K = B_{19} + B_{21}\ ; \quad  A = 2B_{14} - B_{17} - 3B_{19} + 3B_{21} \ ; 
\quad 
B = B_{16} + B_{18} - 2 B_{19} + 2 B_{21}
\end{equation} 
\begin{equation}
C = B_{15} - B_{20} + B_{19} - B_{21} \ ; \quad
D = B_3 + B_4 + B_5 + 3 B_7 +{1\over 2}B_1 +{1\over 3}B_2  
\end{equation} 
\begin{equation}
E = B_6 -{1\over 3}B_4 -{1\over 3}B_5 - B_7 -{1\over 6}B_1 -{1\over 9}B_2 
\ ; \quad
F = B_{14} -{3\over 2}B_4 -{3\over 2}B_5 -{9\over 2}B_7
\end{equation} 
\begin{equation}
G = B_{15} + B_1 +{2\over 3}B_2 + B_4 + 2 B_5 + 3 B_7 \ ; \quad
H = B_{16} -{1\over 2}B_1 -{1\over 3}B_2 - 2 B_4 - B_5 - 3 B_7.
\end{equation}

The general solution of the generic RGE
(\ref{Ki}) for the $O(p^6)$ couplings $K_i$ in terms of the general
solution for the $O(p^4)$ couplings $L_i$ is 
\begin{equation}
K_i(\mu)=K_i(\Lambda)-{1\over
16\pi^2}\left(c_i+b_{ij}L_j(\Lambda)\right)\,log{\mu\over\Lambda}+{1\over 2}{1\over 
(16\pi^2)^2}b_{ij}\Gamma_j\left( log{\mu\over\Lambda}\right)^2.
\end{equation}

There is first of all a renormalization group invariant relation
between the $O(p^6)$ couplings $K_i$ and the $O(p^4)$ couplings $L_i$,
{\it an IR attractive fixed line}. For all ratios $L_i/L_j$ given by their
respective IR fixed points $L_i/L_j=\Gamma_i/\Gamma_j$ the relation is 
\begin{equation}
K_i=(c_i+{1\over 2}b_{ij}L_j){L_m\over\Gamma_m}.
\end{equation}
This predicts the $K_i$ in terms of the $L_i$. For the index $m$ any suitable value
can be chosen, since all the ratios $L_m/\Gamma_m$ are equal in the
fixed point. This is a very interesting and strong prediction. 

Obviously,
the ratios $K_i/K_j$ have IR fixed points for all ratios $L_i/L_j$
sitting in their respective fixed point solutions (\ref{eq:ls})

\begin{equation} 
\left. K_i \over K_j \right|_{{\mathrm f.p.}} ={c_i +(1/2)b_{ik}\Gamma_k\over
c_j + (1/2)b_{jk}\Gamma_k}\ \ \ {\rm for}\ \ \ c_i+(1/2)b_{ik}\Gamma_k,\
c_j+(1/2)b_{jk}\Gamma_k\neq 0. \label{eq:kis}
\end{equation}

All the IR fixed point and fixed line predictions can also be obtained 
within the parameter reduction program.\\

{\bf Acknowledgements.} One of us (M.A.) would like to thank the Institut
f{\"u}r Theoretische Physik und Astrophysik der Universit{\"a}t zu Kiel for
hospitality, and the spanish MEC for a grant of the FPU programme. One 
of us (B.S.) is grateful to the theory group of DESY for continuous hospitality.

\end{document}